 \documentclass[prl,twocolumn,showpacs]{revtex4}

\usepackage{graphicx}
\usepackage{dcolumn}
\usepackage{amsmath}

\begin{document}
\title{Crystal Field and Dzyaloshinsky-Moriya Interaction
in orbitally ordered La$_{0.95}$Sr$_{0.05}$MnO$_3$: An ESR Study}
\author{J.~Deisenhofer $^a$, M.~V.~Eremin $^{a,b}$, D.~V.~Zakharov $^b$,
V.~A.~Ivanshin $^{a,b}$, R.~M.~Eremina $^{a,c}$,
H.-A.~Krug~von~Nidda $^a$, A.~A.~Mukhin $^d$, A.~M.~Balbashov
$^{e}$, and A.~Loidl $^a$}

\affiliation{$^{a\text{ }}$Experimentalphysik V, EKM, Universit\"{a}t
Augsburg, 86135 Augsburg, Germany \\ $^b$ Kazan State University,
420008 Kazan, Russia \\ $^c$ E. K. Zavoisky Physical-Technical
Institute, 420029 Kazan, Russia\\ $^d$ Institut of General
Physics, Russian Academy of Sciences, 117942 Moscow, Russia\\
$^{e}$ Moscow Power Engineering Institute, 105835 Moscow, Russia}
\date{\today}

\begin{abstract}
We present a comprehensive analysis of Dzyaloshinsky-Moriya
interaction and crystal-field parameters using the angular
dependence of the paramagnetic resonance shift and linewidth in
single crystals of La$_{0.95}$Sr$_{0.05}$MnO$_3$ within the
orthorhombic Jahn-Teller distorted phase. The Dzyaloshinsky-Moriya
interaction ($\sim 1$~K) results from the tilting of the MnO$_6$
octahedra against each other. The crystal-field parameters $D$ and
$E$ are found to be of comparable magnitude ($\sim 1$~K) with $D
\approx -E$. This indicates a strong mixing of the
$|3z^2-r^2\rangle$ and $|x^2-y^2\rangle$ states for the real
orbital configuration.

% 76.30.-v Electron paramagnetic resonance and relaxation
% 71.70.Ej Spin-orbit coupling, Zeeman and Stark splitting, Jahn-Teller effect
% 75.30.Et Exchange and superexchange interactions
% 75.30.Vn Colossal magnetoresistance

\end{abstract}

\pacs{76.30.-v, 71.70.Ej, 75.30.Et, 75.30.Vn}

\maketitle

\section{introduction}

The importance of orbital degrees of freedom in understanding the
complex phase diagrams of the manganites \cite{Phasedia} is
subject of intense research activities (see e.g.~references
\cite{Khomskii01,Hill01,Dagotto01} for an overview). The
antiferromagnetic insulator LaMnO$_3$ ($T_{\rm N} = 140$~K) is an
orbitally ordered system \cite{goodenough}, which has been
established experimentally by resonant X-ray scattering
\cite{Murakami98b} and neutron diffraction
\cite{Rodriguez-Carvajal98}. Moreover, Saitoh \textit{et
al.~}recently reported evidence for orbital excitations by Raman
spectroscopy \cite{Saitoh01}. The orbital order in LaMnO$_3$ is
induced by the cooperative Jahn-Teller (JT) effect of the
Mn$^{3+}$ ions (electronic configuration $3d^4:\ t_{2g}^3 e_g^1$,
spin $S = 2$), which at temperatures $T<T_{\rm JT}=750$~K leads to
a strong orthorhombic distortion of the perovskite structure. In
the paramagnetic state electron spin resonance (ESR) reveals a
single exchange-narrowed resonance line with a $g$ value near 2.0
due to all Mn$^{3+}$ ions \cite{Granado98} and hence directly
probes the spin of interest. Doping divalent ions like Sr or Ca
onto the La$^{3+}$ place gradually suppresses the JT distortion
and leads to a ferromagnetic insulating and finally to a metallic
phase at approximately 15\% Sr doping.

In a previous work, we presented a systematic ESR study in single
crystals of La$_{1-x}$Sr$_x$MnO$_3$ with Sr concentrations $0 \leq
x \leq 0.2$ \cite{Ivanshin00}. In the JT distorted phase the
resonance linewidth $\Delta H$ is strongly enhanced compared to
the undistorted phase, reaching maximum values of about $\Delta
H_{\rm max} \approx 2.5$~kOe. Similar results were reported from
polycrystalline La$_{1-x}$Ca$_x$MnO$_3$ \cite{Huber99} and oxygen
doped ceramic LaMnO$_{3+\delta}$ \cite{Tovar99}. Moreover, the
single crystals exhibit a pronounced anisotropy of the resonance
linewidth in the Jahn-Teller distorted phase, which disappears at
temperatures $T > T_{\rm JT}$. The pure LaMnO$_3$ sample turned
out to be strongly twinned and therefore did not allow a detailed
analysis of the angular dependence. Instead, the
La$_{0.95}$Sr$_{0.05}$MnO$_{3}$ crystal was found to be untwinned
and can effectively be treated like the mother compound ($x =
0$), as it is still an antiferromagnetic insulator ($T_{\rm N} =
140$~K) showing a similar magnetic susceptibility. Only the JT
transition is shifted to lower temperatures, with $T_{\rm
JT}(x=0.05) = 600$~K.

In a first approach \cite{Ivanshin00}, we ascribed the orientation
dependence of the ESR linewidth to the influence of the
Dzyaloshinsky-Moriya (DM) interaction, which arises from the
tilting of the MnO$_6$ octahedra along the antiferromagnetically
coupled $b$ axis, only. However, the orthorhombic distortion of
the MnO$_6$ octahedra itself gives rise to a crystal-field (CF)
induced line broadening of comparable order of magnitude
\cite{Huber99} and for a complete description one has also to take
into account the DM interaction via Mn-O-Mn bonds within the
ferromagnetically coupled $ac$ plane, which is smaller than along
the $b$ axis but not negligible.

In the present paper, we study the angular dependence of the
resonance linewidth in the paramagnetic regime at temperatures $T
> T_{\rm N}$. The contributions of CF and DM interaction consist
of the superposition of the four non equivalent Mn ions in the
orthorhombic unit cell. The evaluation is based on the structural
data for LaMnO$_3$ determined from neutron scattering experiments
by Huang \textit{et al.}~\cite{huang}. The application to the
experimental angular dependence of the ESR linewidth at 200K and
300K allows to estimate the microscopic CF parameters $D$ and $E$
and the DM vectors for all Mn-Mn pairs. Throughout this paper we
will use the crystallographic notation following Huang \textit{et
al.}, where the $b$ axis (instead of the former $c$ axis in
reference \cite{Ivanshin00}) denotes the direction perpendicular
to the ferromagnetically coupled $ac$ planes.

\section{Experiment}

The ESR measurements were performed with a Bruker ELEXSYS E500
CW-spectrometer at X-band frequency ($\nu \approx$ 9.35 GHz),
equipped with continuous gas-flow cryostats for He (Oxford
Instruments) and N$_2$ (Bruker), which allow to cover a
temperature range between 4.2 K and 680 K. The ESR spectra record
the power $P$ absorbed by the sample from the transverse magnetic
microwave field as a function of the static magnetic field $H$.
The signal-to-noise ratio of the spectra is improved by detecting
the derivative $dP/dH$ with lock-in technique. A small single
crystal of La$_{0.95}$Sr$_{0.05}$MnO$_3$ (volume 2~mm$^3$) was
fixed on a suprasil quartz rod with low-temperature glue (General
Electrics), which allowed the rotation of the sample around
defined crystallographic axes.
\begin{figure}[h]
\centering
\includegraphics[width=60mm,clip,angle=-90]{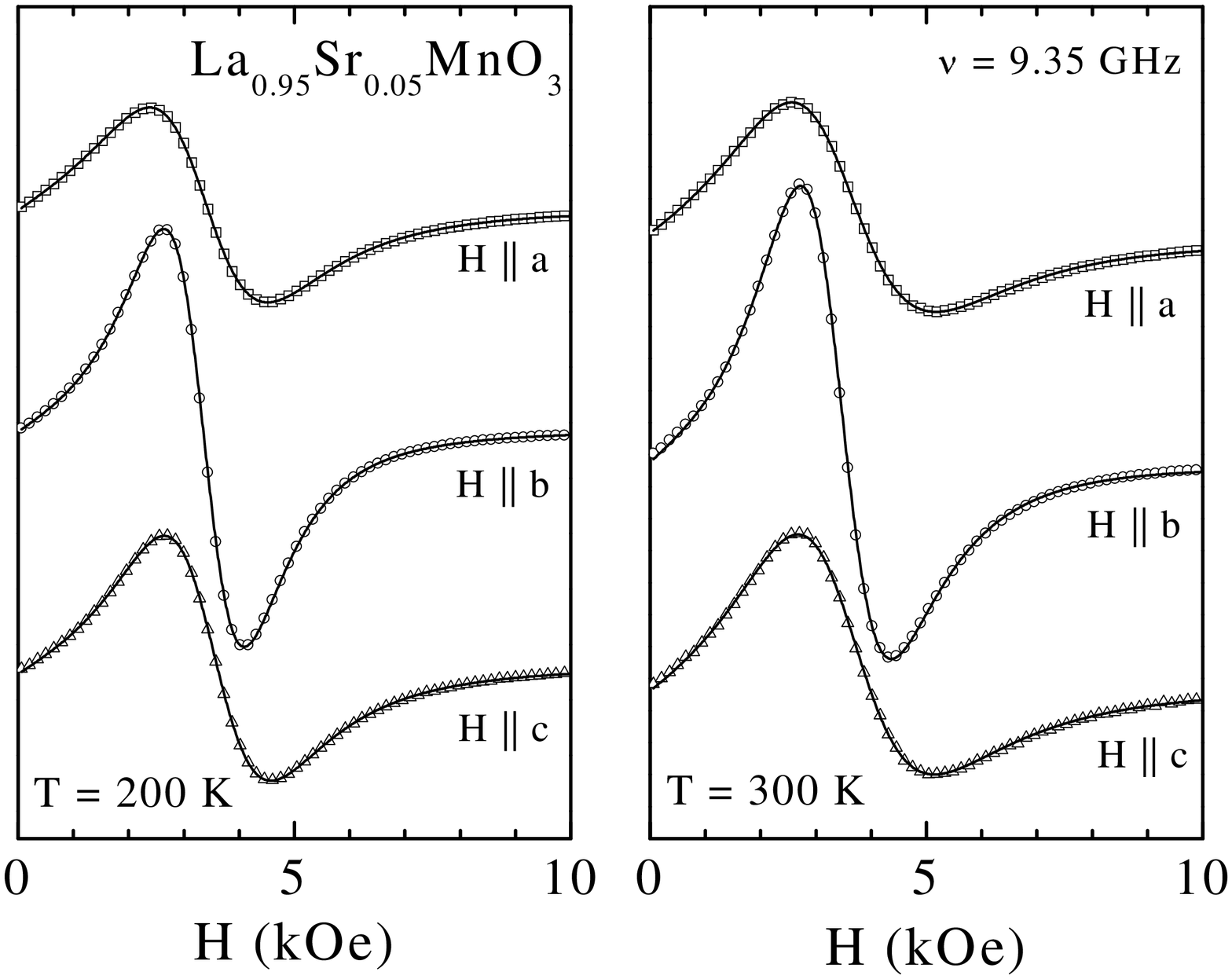}
\vspace{2mm} \caption{ESR spectra of La$_{0.95}$Sr$_{0.05}$MnO$_3$
at 200~K (left hand side) and 300~K (right hand side) for the
magnetic field $H$ applied parallel to the crystallographic axis
($a$,$b$,$c$). Solid lines represent the fits using the Dysonian
line shape, equation \ref{dyson}.} \label{spectra}
\end{figure}
We measured the angular dependence of the paramagnetic resonance
with respect to the orientation of the static magnetic field $H$
in the three crystallographic planes ($ab$, $ac$, and $bc$) at
temperatures 200 K and 300 K, deeply in the Jahn-Teller distorted
orthorhombic phase. Figure \ref{spectra} illustrates typical ESR
spectra for the magnetic field applied parallel to the three main
axes. In all cases, one observes a broad, exchange-narrowed
resonance line, which is well fitted by a Dysonian line shape
\cite{barnes} given by

\begin{equation}
\frac{dP}{dH} \propto \frac{d}{dH}\{\frac{\Delta H + \alpha
(H-H_{\rm res})}{(H-H_{\rm res})^2 + \Delta H^2} + \frac{\Delta H
+ \alpha (H+H_{\rm res})}{(H+H_{\rm res})^2 + \Delta H^2}\}
\label{dyson}
\end{equation}

This is an asymmetric Lorentzian line, which includes both
absorption and dispersion, where $\alpha$ denotes the
dispersion-to-absorption (D/A) ratio. As the linewidth $\Delta H$
is of the same order of magnitude as the resonance field $H_{\rm
res}$ in the present compounds, eq.~(\ref{dyson}) takes into
account both circular components of the exciting linearly
polarized microwave field and therefore also includes the
resonance at reversed magnetic field $-H_{\rm res}$.

Such asymmetric line shapes are usually observed in metals, where
the skin effect drives electric and magnetic microwave components
in the sample out of phase and therefore leads to an admixture of
dispersion into the absorption spectra. For samples small compared
to the skin depth one expects a symmetric absorption spectrum
($\alpha$ = 0), whereas for samples large compared to the skin
depth absorption and dispersion are of equal strength yielding an
asymmetric resonance line ($\alpha$ = 1). A second reason for the
asymmetry, which also occurs in insulators, arises from the
influence of non diagonal elements of the dynamic susceptibility:
In systems with interactions of low symmetry and sufficiently
broad resonance lines ($H_{\rm res} \approx \Delta H$) the line
shape shows characteristic distortions depending on the frequency
and orientation of the exciting microwave field \cite{benner},
where eq.~(\ref{dyson}) yields a useful approximation of the
spectrum.

Comparing the spectra at 200 K with those at 300 K one recognizes
an increasing asymmetry with increasing temperature: The average
D/A ratio increases from about 0.05 at 200 K to 0.35 at 300 K.
This can be understood in terms of the skin effect due to the
increase of the conductivity with increasing temperature. The
resistivity $\rho$ of the sample under investigation is about
$500$~$\Omega$cm at 200~K and $10$~$\Omega$cm at 300~K
\cite{mukhin}. We estimated \cite{Ivanshin00} that for
resistivities $\rho < 1$~$\Omega$cm the skin depth becomes smaller
than the dimensions of the sample, which are about 1~mm. As the
skin depth is proportional to the square root of the resistivity,
the influence of the skin effect becomes visible even at 300~K.
However at 200~K, it can be neglected. The remaining asymmetry,
which will be discussed in connection with the resonance field
below, has to be ascribed to the non diagonal contributions of the
dynamic susceptibility.

\begin{figure}[h]
\centering
\includegraphics[width=80mm,clip]{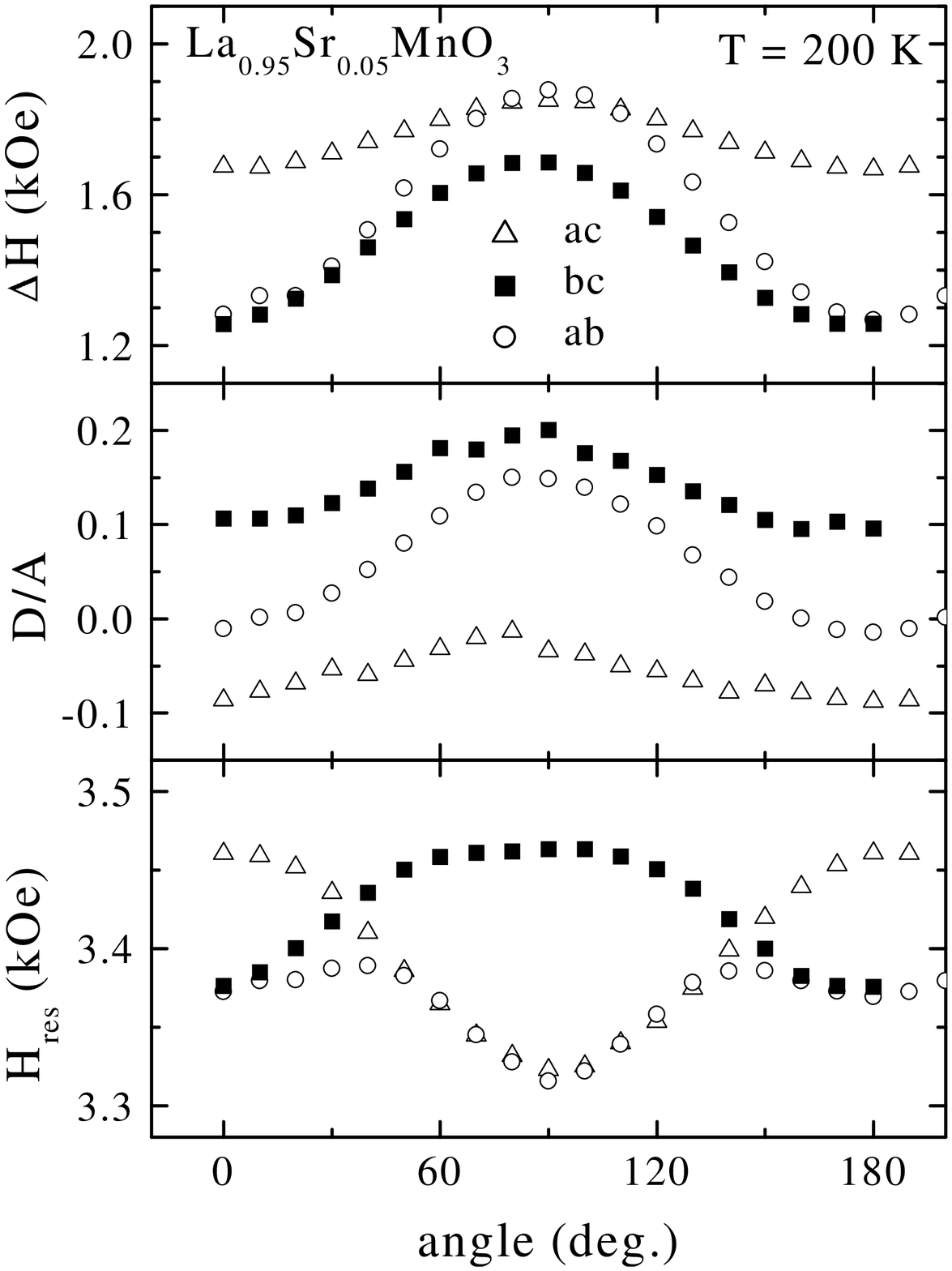}
\vspace{2mm} \caption{Angular dependence of linewidth $\Delta H$
(upper frame), D/A ratio (middle frame), and resonance field
$H_{\rm res}$ (lower frame) in La$_{0.95}$Sr$_{0.05}$MnO$_3$ for
the magnetic field applied within the three crystallographic
planes at 200~K} \label{exp200K}
\end{figure}
\begin{figure}[h]
\centering
\includegraphics[width=80mm,clip]{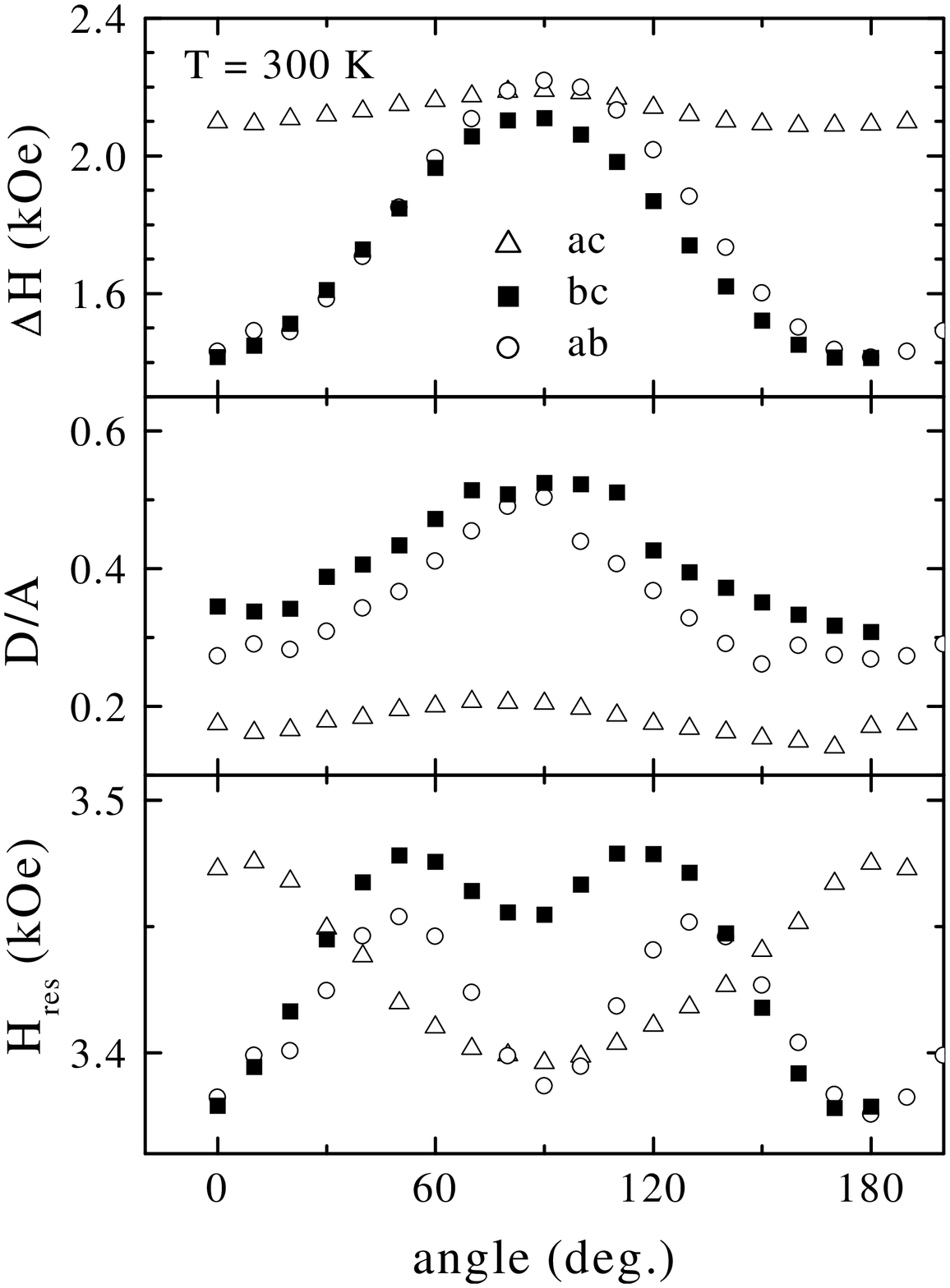}
\vspace{2mm} \caption{Angular dependence of linewidth $\Delta H$
(upper frame), D/A ratio (middle frame), and resonance field
$H_{\rm res}$ (lower frame) in La$_{0.95}$Sr$_{0.05}$MnO$_3$ for
the magnetic field applied within the three crystallographic
planes at 300~K} \label{exp300K}
\end{figure}

Figures \ref{exp200K} and \ref{exp300K} show the full angular
dependence of linewidth (upper frame), resonance field (lower
frame) and D/A ratio (middle frame) for 200~K and 300~K, as
obtained from the fit with eq.~\ref{dyson}. The linewidth
exhibits a pronounced anisotropy with respect to the
crystallographic $b$ axis and a weaker angular dependence within
the $ac$ plane, which can be empirically described by a
$\cos^2$-law. The D/A ratio depends on the orientation of the
microwave field, which is applied parallel to the rotation axis,
and even becomes negative at 200~K, if the microwave field is
perpendicular to the $ac$ plane. This cannot be understood in
terms of the skin effect alone, which always produces a positive
D/A ratio. Moreover, the influence of the skin effect should be
negligible at 200~K. Hence, the asymmetry has to be ascribed
partly to the influence of the non diagonal elements of the
dynamic susceptibility.
The angular dependence of the resonance field exhibits additional
$90^{\circ}$ modulations. They appear to be strongest for the $ab$
plane, where the D/A ratio exhibits the most pronounced angular
dependence. At the same time the D/A ratio is strongly correlated
with the resonance linewidth, as they both attain their maximum
values at the same angle. However, for these broad signals, the
baseline is not so well defined as in the case of narrow lines and
a shift in the D/A ratio can be compensated by resonance field and
baseline. For this reason, the angular dependence of the D/A ratio
can be an artificial effect, which is probably related to the
large linewidth. Hence, we tried to fit the ESR spectra with a
constant average D/A ratio for each rotation plane. It turned out
that the spectra at 200~K can be satisfactorily fitted as well.
This procedure has no visible influence on the linewidth, but the
resonance field is clearly changed, as can be obtained from
Fig.~\ref{r200Kcor}, which shows the corrected data for 200~K. The
amplitude of the anisotropy remains approximately unchanged, but
the additional $90^{\circ}$ modulations more or less disappear.
Turning to the resonance field data at 300~K, the situation is
more complicated, as the influence of the skin effect becomes
important, too. The amplitude of the D/A-angular dependence
increases and a fit with a constant D/A ratio is not satisfactory
anymore. Therefore we omit a correction of the resonance fields at
300~K.

\section{Theoretical background}

A strongly exchange coupled magnetic system like LaMnO$_3$ can be
described by the following Hamiltonian

\begin{equation}\label{hamiltonian}
{\cal H} = J\sum\limits_{(i<j)}{\mathbf S}_{i}\cdot {\mathbf
S}_{j}-\mu_{\rm B}\sum\limits_{i}{\mathbf H}\cdot {\mathbf
g}\cdot {\mathbf S}_{i}+{\cal H_{\rm int}}
\end{equation}

where the first term describes the the superexchange interaction
between two next-neighbor Mn spins ${\bf S}_i$ and ${\bf S}_j$
with coupling constant $J$. The second term describes the Zeeman
splitting of the spin states with gyromagnetic tensor $\mathbf{g}$
within an external magnetic field ${\bf H}$, where $\mu_{\rm B}$
denotes the Bohr magneton. The third term ${\cal H}_{\rm int}$
includes all interactions, which do not conserve the total spin
and therefore contribute to the broadening of the ESR line as
there are CF, DM interaction, anisotropic exchange (AE)
interaction, dipole-dipole interaction, and hyper-fine
interaction. Due to estimation of their relative strength
\cite{Huber99} it turned out that CF and DM interaction yield by
far the largest contribution.

\subsection{Interactions}
To derive the appropriate expressions for CF and DM interaction,
we use the structural parameters of the LaMnO$_3$-IIa sample
(space group \textit{Pnma}) determined by Huang \textit{et
al.}~\cite{huang}, because it shows an antiferromagnetic ground
state and an ordering temperature of about 140 K, which is
consistent with the magnetic properties of our sample
\cite{paras}. Figure 7(a) in reference \cite{huang} shows the
crystallographic structure of LaMnO$_3$ in the strongly
Jahn-Teller distorted phase. Due to the tilting of the MnO$_6$
octahedra one can identify four inequivalent positions of
manganese ions in the orthorhombic unit cell, which is
illustrated in Fig. 7(b) of the same reference \cite{huang}.

\subsubsection{Crystal Field}
In a local coordinate system, where the axes are directed along
the Mn-O bonds of the MnO$_{6}$ octahedra, the spin Hamiltonian
is usually written in cartesian spin components ($S_x,S_y,S_z$)
with parameters $D$ and $E$ as \cite{abragam}

\begin{equation}
{\cal H^{\rm CF}} =D S_{z}^{2}+E (S_{x}^{2}-S_{y}^{2})
\end{equation}

where the $z$ axis is directed along the longest Mn-O link close
to the $ac$ plane and the $y$ axis is parallel to Mn-O(1) bond
almost along the $b$ direction. Small orthorhombic distortions of
the MnO$_{6}$ octahedron will be neglected in this section.
Because the linewidth is invariant under shift transformations, we
can redefine ${\cal H^{\rm CF}}$ for simplification as

\begin{eqnarray}\label{CFham}
{\cal H}^{\rm CF} &=& DS_{z}^{2}+E(S_{x}^{2}-S_{y}^{2})-E
(S_{z}^{2}+S_{x}^{2}+S_{y}^{2}) \nonumber \\ &=&
D^{\prime}S_{z}^{2}+E^{\prime}S_{y}^{2}
\end{eqnarray}

where $D^{\prime}=D-E$ and $E^{\prime} = -2E$. It is useful to
recall that the $y$ axis is almost parallel to $b$. The octahedra
are rotated with respect to their next neighbor within the unit
cell. Therefore in the crystallographic system ($a,b,c$) we have
different crystal-field components at the four inequivalent Mn
places which are listed in table \ref{CFpara}.

The CF components for the Mn ion at $(0,0,\frac{1}{2})$ read

\begin{eqnarray}\label{CFcomp}
D_{xx}^{(1)}&=&(\frac{1}{2}-x)^{2}(\frac{a}{R_{l}})^{2}D^{\prime}+
(\frac{1}{2}-X)^{2}(\frac{a}{R_{m}})^{2}E^{\prime}
\\ \nonumber
D_{yy}^{(1)}&=&(\frac{yb}{R_{l}})^{2}D^{\prime}+(\frac{b}{4R_{m}})^{2}E^{\prime}
\\ \nonumber
D_{zz}^{(1)}&=&(\frac{zc}{R_{l}})^{2}D^{\prime}+(\frac{Zc}{R_{m}})^{2}E^{\prime}
\\ \nonumber
D_{xy}^{(1)}&=&(x-\frac{1}{2})\frac{ayb}{R_{l}^{2}}D^{\prime}+(X-\frac{1}{2})\frac{ab}{4R_{m}^{2}}E^{\prime}
\\ \nonumber
D_{xz}^{(1)}&=&(\frac{1}{2}-x)\frac{azc}{R_{l}^{2}}D^{\prime}+(\frac{1}{2}-X)\frac{aZc}{R_{m}^{2}}E^{\prime}
\\ \nonumber
D_{yz}^{(1)}&=&-\frac{ybzc}{R_{l}^{2}}D^{\prime}-\frac{bZc}{4R_{m}^{2}}E^{\prime}
\end{eqnarray}

Long, middle and short Mn-O distances in square are

\begin{eqnarray}\label{MnObond}
R_{l}^{2}&=&(x-\frac{1}{2})^{2}a^{2}+(yb)^{2}+(zc)^{2} \\
\nonumber
R_{m}^{2}&=&(X-\frac{1}{2})^{2}a^{2}+(\frac{b}{4})^{2}+(Zc)^{2} \\
\nonumber
R_{s}^{2}&=&(xa)^{2}+(yb)^{2}+(\frac{1}{2}+z)^{2}c^{2}
\end{eqnarray}

Here capital $X$, $Z$ and small letters $x$, $y$, $z$ denote the
structure parameters (after Huang \textit{et al.} \cite{huang})
for the oxygen ions along $b$ and within the ac plane
respectively.

\subsubsection{Dzyaloshinsky-Moriya Interaction}
\begin{figure}[h]
\centering
\includegraphics[width=70mm,clip]{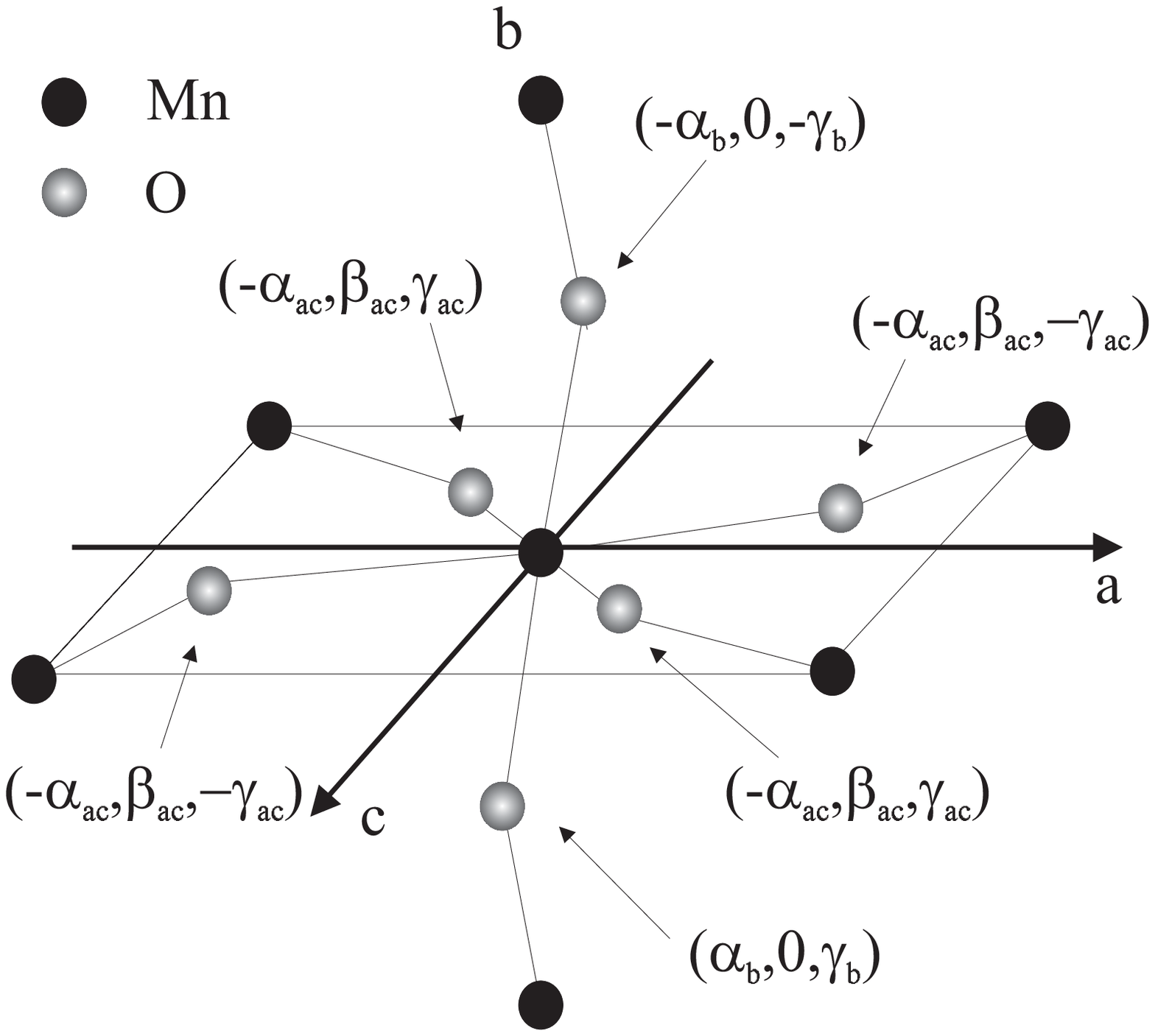}
\vspace{2mm} \caption{Next-neighbor bonds of the Mn ions.
$\alpha$, $\beta$, and $\gamma$ denote the cartesian components of
the DM vector ${\bf G}_{ij}$.} \label{dmchain}
\end{figure}

The Hamiltonian, which describes the antisymmetric DM interaction
\cite{moriya}, can be written as

\begin{equation}\label{dmham}
{\cal H}_{DM} = {\mathbf G}_{ij}\cdot[{\mathbf S}_{i} \times
{\mathbf S}_{j}],
\end{equation}

with the DM vector ${\mathbf G}_{ij} = d_{ij} \cdot[{\bf n}_{Oi}
\times {\bf n}_{Oj}]$ being perpendicular to the plane defined by
a Mn ion at site ($i$), the bridge ligand O, and a Mn ion at site
($j$), where ${\bf n}$ are unit vectors along Mn$^{3+}$-O$^{2-}$
bonds \cite{keffer,moskvin}. The intrinsic scalar parameter
$d_{ij}$ strongly depends on the orbital states and the Mn-O-Mn
bridge angle. In the case of pure LaMnO$_3$ both the tilting and
the JT distortion of the MnO$_6$ octahedra account for the origin
of antisymmetric contributions to the superexchange interaction
between the Mn ions. A necessary condition for the existence of DM
contributions is the lack of a center of inversion between the
magnetic ions \cite{moriya}. With the apical oxygen being shifted
away from the [010] axis, there is a rather strong DM coupling
between the $ac$ planes and a smaller coupling within the $ac$
planes. Figure \ref{dmchain} depicts all next-neighbor couplings
that give rise to a DM interaction.
For the \textit{Pnma} structure, the components
${G}_{ij}^{\alpha}$ ($\alpha =x,y,z$) of the DM vectors of all Mn
pairs within the unit cell are listed in table \ref{DMpara}. The
index $j = 1...4$ denotes the four magnetically inequivalent
positions of the Mn ion in the unit cell. The index $i = 1...6$
refers to the six next neighbors around each Mn site with number
$j$, as it is shown in Fig. \ref{dmchain}. The absolute values
can be expressed via two parameters $d_{1}$ (inter $ac$ plane)
and $d_{2}$ (intra $ac$ plane) as

\begin{eqnarray}
\alpha_{b}&=& d_{1}\frac{Z}{2}\frac{bc}{R_{m}^{2}} \\ \nonumber
\beta_{b}&=&0 \\ \nonumber
\gamma_{b}&=&d_{1}\frac{1}{2}(X-\frac{1}{2})\frac{ab}{R_{m}^{2}}
\\ \nonumber
\alpha_{ac}&=&d_{2}\frac{y}{2}\frac{bc}{R_{l}R_{s}} \\ \nonumber
\beta_{ac}&=&d_{2}\frac{1}{2}(x-z-\frac{1}{2})\frac{ac}{R_{l}R_{s}}
\\ \nonumber
\gamma_{ac}&=&d_{2}\frac{y}{2}\frac{ab}{R_{l}R_{s}}
\end{eqnarray}

where $R_m$ measures the Mn-O distance along the
antiferromagnetically coupled $b$ direction and $R_l$ and $R_s$
denote the Mn-O distances within the ferromagnetically coupled
$ac$-plane. A microscopic expression for $d_{1}$ and $d_{2}$ is
discussed in Appendix 1.

\subsection{Resonance Field}

The resonance field of the strongly exchange narrowed ESR line is
generally determined by the first moment of the spectrum. Here we
choose another approach, which also allows to take into account
the demagnetization effect. Hence, we start from the equation of
motion \cite{moreno}:

\begin{eqnarray}\label{eqmotion}
\frac{\partial S^{+}}{\partial t}&=&ig\mu_{\rm B}\{H-\frac{M}{2}
(2N_{zz}-N_{xx}-N_{yy})\}S^{+}\nonumber \\
&+&\frac{1}{i\hbar}[S^{+},H_{\rm int}]+ig\mu_{\rm
B}M(N_{xx}-N_{yy})S^{-}
\end{eqnarray}

with the magnetization $M$ and the demagnetizing factors
$N_{\alpha \alpha}$. The circular spin operators are defined as
usual $S^{\pm} = S_x \pm iS_y$. Here we assume an isotropic $g$ value,
because the influence of the AE interaction can be neglected with
respect to the crystal field. The DM interaction does not
contribute to the resonance shift, because the expectation value
of the commutator $\langle[S^{+},H^{\rm DM}]\rangle$ vanishes.
After linearization the expectation value of the commutator
$\langle[S^{+},H_{\rm int}]\rangle$ is given by

\begin{eqnarray}\label{CFcommutator}
\langle[S^{+},H^{\rm CF}]\rangle &\approx& \frac{M_{\rm
at}}{g\mu_{\rm
B}}\frac{1}{4}\sum\limits_{j}(\tilde{D}_{xx}^{(j)}-\tilde{D}_{yy}^{(j)}
+2i\tilde{D}_{xy}^{(j)})S^{-} \nonumber \\ &+& \frac{M_{\rm
at}}{g\mu_{\rm
B}}\frac{1}{4}\sum\limits_{j}(\tilde{D}_{xx}^{(j)}+\tilde{D}_{yy}^{(j)}
-2\tilde{D}_{zz}^{(j)})S^{+}
\end{eqnarray}

Here $M_{\rm at}=(g\mu_{\rm B})^{2}S(S+1)H/[3k(T-\Theta_{\rm
CW})]$ is the magnetization per one Mn site, and $\Theta_{\rm CW}
= 111$~K is the paramagnetic Curie-Weiss temperature. The
crystal-field components $\tilde{D}_{\alpha \beta}$ refer to the
coordinate system, in which the external field determines the $z$
axis and are related to the components $D_{\alpha \beta}$ in the
crystallographic system via the usual transformation rules (see
Appendix 2). The sum ($j$) is running over the four inequivalent
Mn positions of the unit cell. If we further neglect the
demagnetization effect, which according to \cite{moreno} is
relatively small in the paramagnetic regime, the ESR frequency
reads

\begin{eqnarray}\label{resonance}
(h\nu )^{2}&=&[g\mu_{\rm B}H+\frac{M_{\rm at}}{4g\mu_{\rm
B}}\sum\limits_{j}
(\tilde{D}_{xx}^{(j)}+\tilde{D}_{yy}^{(j)}-2\tilde{D}_{zz}^{(j)})]^{2}
\\ \nonumber
&-&{\frac{M_{\rm at}}{4g\mu_{\rm B}}}^2[\sum\limits_{j}
(\tilde{D}_{xx}^{(j)}-\tilde{D}_{yy}^{(j)}+2i\tilde{D}_{xy}^{(j)})]
\\ \nonumber &\times& [\sum\limits_{j}
(\tilde{D}_{xx}^{(j)}-\tilde{D}_{yy}^{(j)}-2i\tilde{D}_{xy}^{(j)})]
\end{eqnarray}

The angular dependence of the resonance field $H_{\rm res}
(\theta,\varphi)$ is obtained after substitution of
$\tilde{D}_{\alpha \beta}$ by $D_{\alpha \beta}$ (see Appendix 2).
Polar angle $\theta$ and azimuth angle $\varphi$ are measured
with respect to the $c$ and $a$ axes of the orthorhobmic unit
cell.

\subsection{Linewidth}

In the case of strong exchange narrowing $({\cal H}_{\rm ex} \gg
{\cal H}_{\rm int})$, the ESR linewidth $\Delta H$ is determined
by the second moment $M_2$ of the resonance line divided by the
exchange frequency $\omega_{\rm ex}$ \cite{anderson}

\begin{equation}
\Delta H \simeq \frac{1}{g \mu _{\rm B} \hbar}
\frac{M_2}{\omega_{\rm ex}} \label{narrowed}
\end{equation}

In the present case the second moment is determined by the sum of
DM and CF contributions $(M_{2}^{\rm DM}+M_{2}^{\rm CF})$

Within the coordinate system, where the z axis is determined by
the external magnetic field, the second moment due to
Dzyaloshinsky-Moriya interaction is calculated as
\cite{castner,soos}

\begin{eqnarray}\label{M2DM}
{M_{2}^{DM}=\frac{2}{3}S(S+1)\sum\limits_{i,j}\left[(\tilde{G}_{ij}^{x})^{2}+
(\tilde{G}_{ij}^{y})^{2}+2(\tilde{G}_{ij}^{z})^{2}\right]}
\end{eqnarray}

The index $j = 1...4$ is running over all four magnetically
inequivalent positions of the Mn ion in the unit cell. The sum
over index $i = 1...6$ refers to the six next Mn neighbors around
each Mn site with number $j$. After transformation into the
crystallographic system and using the values listed in table
\ref{DMpara}, the average over all four positions yields the
angular dependence

\begin{eqnarray}\label{M2DM2}
M_{2}^{\rm DM} &=& \frac{2}{3}S(S+1)\{
(2\alpha_{b}^{2}+4\alpha_{ab}^{2})[1+\sin^{2}\theta\cos^{2}\varphi]
\nonumber \\
&+& 4\beta _{ab}^{2}[1+\sin^{2}\theta\sin^{2}\varphi]\nonumber \\
&+&(2\gamma_{b}^{2}+4\gamma_{ab}^{2})[1+\cos^{2}\theta]\}
\end{eqnarray}

For the crystal field, the expression for the second moment is
given in terms of $\tilde{D}_{\alpha \beta }$ by

\begin{eqnarray}\label{M2CF}
M_{2}^{CF}&=&\frac{4S(S+1)-3}{20 \cdot 4} \sum\limits_{j}\{f_{1}[2
\tilde{D}_{zz}^{(j)}-\tilde{D}_{xx}^{(j)}-\tilde{D}_{yy}^{(j)}]^{2}
\nonumber \\ &+& 10f_{2}(\tilde{D}_{xz}^{2}+\tilde{D}_{yz}^{2})\nonumber \\
&+& f_{3}[(\tilde{D}_{xx}
-\tilde{D}_{yy})^{2}+4\tilde{D}_{xy}^{2}]\}
\end{eqnarray}

and again the sum over $j$ is running over the four inequivalent
Mn positions in the unit cell. The prefactors $f_{k}$ allow to
separate secular ($f_1$) and non secular ($f_2$ and $f_3$)
contributions. Again, the angular dependence is obtained by
transforming the CF components into the crystallographic system
and using the values listed in table \ref{CFpara}. The full
expressions divided in secular and nonsecular parts are given in
Appendix 2. It is important to point out here that both secular
and nonsecular parts contain strong contributions proportional to
$\cos 4 \varphi$ and $\sin 4\theta$, which should result in a
$\pi/2$ periodic modulation of the angular dependence. However, in
the sum $M_{2}^{\rm CF}({\rm secular})+M_{2}^{\rm CF}({\rm
nonsecular})$ these contributions cancel each other for the case
$f_{1}$ = $f_{2}$ = $f_{3}$. As the experimental data do not
exhibit any $\pi/2$ periodic modulation, only the latter case
yields the appropriate description of the observed angular
dependence. Under these circumstances the crystal field
contribution reads e.g. for the $ab$ plane

\begin{eqnarray}
M_{2}^{\rm CF}(ab)&=&\frac{1}{80}[4S(S+1)-3]\{
\frac{1}{2}\sum\limits_j [2D_{zz}^{(j)}-D_{xx}^{(j)}-D_{yy}^{(j)}]^{2} \nonumber \\
&+&\frac{5}{2}\sum\limits_j[(D_{xx}^{(j)}-D_{yy}^{(j)})^{2}+4(D_{xy}^{(j)})^{2}]
\nonumber \\
&+& 7\sum\limits_j\left[(D_{zy}^{(j)})^{2}+(D_{xz}^{(j)})^{2}\right] \\
&+&\sum\limits_j\left[3(D_{zy}^{(j)})^{2}-3(D_{xz}^{(j)})^{2}\right.\nonumber\\
&-& \left.(2D_{zz}^{(j)}-D_{xx}^{(j)}-D_{yy}^{(j)})
(D_{xx}^{(j)}-D_{yy}^{(j)})\right ]\cos2\varphi \}\nonumber
\end{eqnarray}

The respective expressions for $ac$ and $bc$ plane are obtained
by permutation of $(x,y,z)$ and exchange of
$\varphi\rightarrow\theta$, as indicated in Appendix 2.

\section{ANALYSIS and DISCUSSION}

\begin{figure}[h]
\centering
\includegraphics[width=80mm,clip]{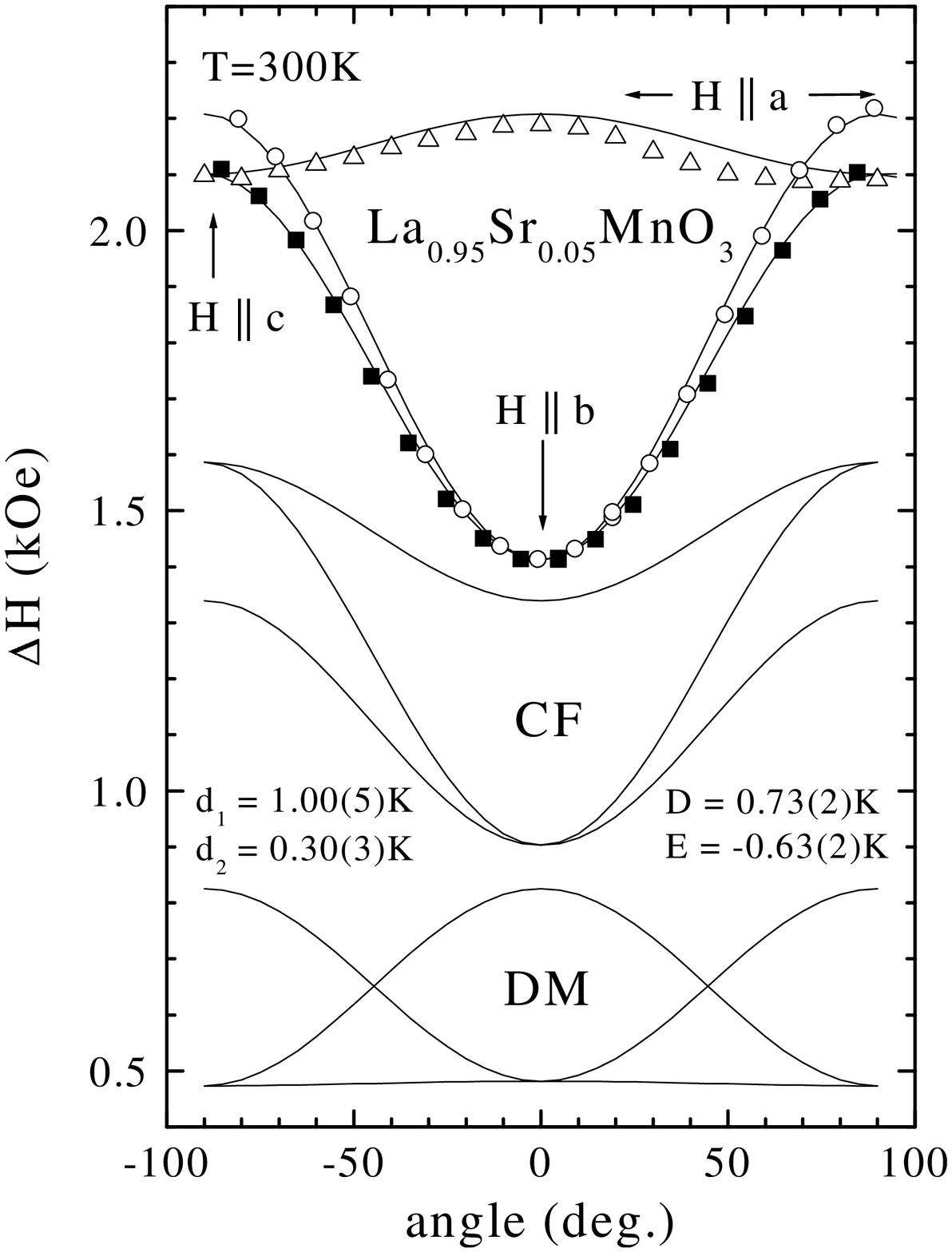}
\vspace{2mm} \caption{Angular dependence of the ESR linewidth
$\Delta H$ in La$_{0.95}$Sr$_{0.05}$MnO$_3$ for the magnetic field
applied within the three crystallographic planes at 300 K. The
solid lines represent the fit with equation \ref{narrowed}. The
lines below illustrate the contributions of CF and DM interaction,
respectively} \label{linewidth}
\end{figure}

As it is shown in Fig.~\ref{linewidth}, the linewidth data at
300~K are well described by the superposition of CF and DM
contribution to the second moment (Eq.~\ref{narrowed}).

The superexchange integral $J$ was estimated in mean-field
approximation from the Curie-Weiss temperature $\Theta_{\rm CW} =
C(4 J_{ac} + 2 J_b)$ with ferromagnetic in-plane coupling $J_{ac}$
to four neighboring Mn ions and antiferromagnetic inter-plane
coupling $J_b$ to two neighboring Mn ions and the Curie constant
$C=S(S+1)/3k_{\rm B}$ ($k_{\rm B}$: Boltzmann constant). Assuming
$-J_{ac} = J_b = J$ and inserting the Mn$^{3+}$ spin $S=2$ and
$\Theta_{\rm CW} = 111$~K \cite{paras}, we obtain $J=14$~K. Both
DM and CF interaction are of equal order of magnitude about 1K.
The DM interaction $d_1=1.00(5)$ K along the antiferromagnetically
coupled $b$ axis is about three times larger than $d_2=0.30(3)$ K
within the ferromagnetically coupled $ac$ plane. The absolute
values of the crystal field parameters $D=0.73(2)$ K and
$E=-0.63(2)$ K are of nearly equal strength indicating comparable
axial ($D$) and rhombic ($E$) distortions of the MnO$_6$
octahedra.

\begin{figure}[h]
\centering
\includegraphics[width=80mm,clip]{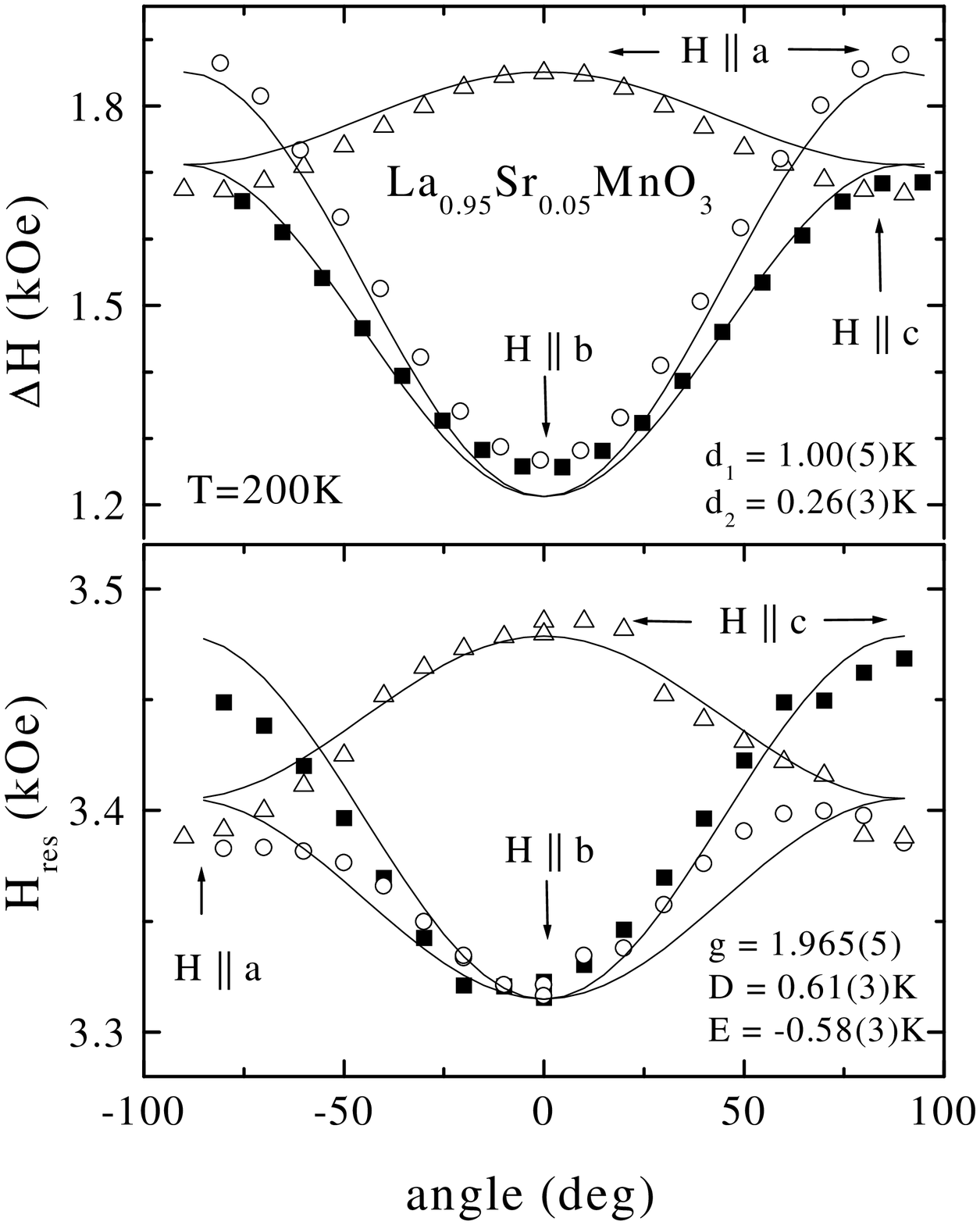}
\vspace{2mm} \caption{Angular dependence of the ESR linewidth
$\Delta H$ (upper frame) and corrected resonance field $H_{\rm
res}$ (lower frame) in La$_{0.95}$Sr$_{0.05}$MnO$_3$ for the
magnetic field applied within the three crystallographic planes at
200~K. The solid lines represent the fit with equations
\ref{narrowed} and \ref{resonance}, respectively.}
\label{r200Kcor}
\end{figure}

The angular dependence of the resonance linewidth at 200~K is well
fitted with similar but slightly smaller parameters ($d_1=1.00(5)$
K, $d_1=0.26(3)$ K, $D=0.61(3)$ K, $E=-0.58(3)$ K) in
Fig.~\ref{r200Kcor}. In addition, the resonance field ($g=1.965$)
can be described by similar crystal field parameters as the
linewidth, thereby showing good agreement of static and dynamic
susceptibility.
As the derivation of the linewidth was carried out in the
high-temperature limit ($T \gg \Theta_{\rm CW}$), where it
approximates an asymptotic value $\Delta H (\infty)$, we have to
take into account, that the fit parameters $D$, $E$, $d_1$ and
$d_2$ should reflect the temperature dependence. According to
Huber \textit{et al.}~\cite{Huber99} far apart from any magnetic
or structural transitions, the temperature dependence can be
approximated by the quotient of the single-ion Curie
susceptibility and the experimental Curie-Weiss susceptibility as

\begin{equation}\label{huber}
\Delta H (T) = \frac{T-\Theta_{\rm CW}}{T} \Delta H(\infty).
\end{equation}

Taking into account that the square of CF and DM parameters
appears in the formulae for the second moment, we can extrapolate
these parameters to $T \rightarrow \infty$ via the square root of
Eq.~\ref{huber}. Using the data at 300~K, which is far above
magnetic order, we have to multiply the parameters by a factor
1.26 and get $D(\infty) = 0.91$~K, $E(\infty) = -0.79$~K,
$d_1(\infty) = 1.26$~K and $d_2(\infty) = 0.38$~K. Comparing the
values at 200~K, one expects a reduction of the parameters by
approximately $15\%$ with respect to 300~K. Indeed, this is well
fulfilled, despite the onset of the critical behavior on
approaching magnetic order at 140~K, only $d_1$ remains unchanged.

The equations for the resonance field already contain the
temperature dependence within the magnetization $M(T)$. For this
reason it should be better described by the high-temperature
values of the CF parameters than by the temperature dependent
ones, which is the case at 200 K. Due to the rather large
uncertainty in the determination of the resonance field and the
proximity of the critical temperature region, this discrepancy is
not assumed to be of significant importance. Thus, at 300~K the
amplitude of the uncorrected resonance-field anisotropy is nicely
reproduced by the high-temperature values of the CF parameters as
$H_{\rm res}^{\rm max} - H_{\rm res}^{\rm min} = 95$~Oe.

The strength of CF and DM interaction was independently determined
from orientation dependent magnetization data \cite{Pimenov00},
measured at 4.2~K in a La$_{0.95}$Sr$_{0.05}$MnO$_3$ single
crystal of the same batch as used in our ESR experiments. In
combination with antiferromagnetic resonance measurements it was
shown there, that the ground state ($T < T_{\rm N}$) exhibits a
canted antiferromagnetic structure, which rules out any phase
separation in ferromagnetic and antiferromagnetic regions. Within
the two-sublattice model Pimenov \textit{et al.}~obtained (after
translation to our Hamiltonian in units of Kelvin) to the
following parameters from the magnetization data \cite{Pimenov00}:
$J=13.2$~K, $K_z = -2E = 0.93$~K, $K_x = D-E = 0.95$~K, and $G =
sin(155^{\circ}) \dot d_1 = 1.18$~K. These results are in good
agreement with our findings, as the values of the CF parameters
$D = 0.48$~K and $E = - 0.47$~K are essentially equal. The DM
contribution is of similar strength, where $d_2$ was neglected.
Thus, we could confirm the validity of their model from a
microscopic point of view.

ESR studies by Huber \textit{et al.}~in LaMnO$_3$ \cite{Huber99}
and Tovar \textit{et al.}~\cite{Tovar99} in the series
LaMnO$_{3+\delta}$ revealed values comparable to our parameters.
For pure LaMnO$_3$, they use the CF parameter $D = 1.92$~K
determined by Moussa \textit{et al.}~from neutron-scattering
experiments \cite{moussa} and estimate the DM contribution as $d
\approx 0.8$~K. For an oxygen excess $\delta = 0.03$ with a
transition temperature $T_{\rm JT} = 600$~K comparable to
La$_{0.95}$Sr$_{0.05}$MnO$_3$, the CF parameter $D$ is found to
be about $20\%$ smaller than in the pure compound, whereas the DM
interaction remains essentially unchanged at $d \approx 0.8$~K.
Though these data were measured in polycrystalline samples, where
no distinction between axial($D$) and rhombic($E$) CF parameters
and the inter-plane and intra-plane DM interaction could be made,
their results agree with ours within a factor of 2.

The fact, that the two CF parameter are of comparable strength is
of particular importance for the picture of orbital order in
LaMnO$_3$. Many illustrations of the orbitally ordered state
invoke the ordering of $d_{3z^2-r^2}$ orbitals
(e.g.~\cite{Saitoh01,Solovyev96,Murakami98b}), as originally
supposed by Goodenough \cite{goodenough} in order to describe the
antiferromagnetic superexchange. So far this picture has been
widely applied and in the absence of an experimentum crucis, which
directly probes not only the electron-electron interaction induced
by orbital order (which resonant X-Ray scattering is assumed to be
able to \cite{Murakami98b}) but also reveals the kind of orbitals
involved (which could not be identified by X-Ray resonant
scattering \cite{Ishihara98b}), it is the obvious choice in view
of the observed axial distortion of the MnO$_6$ octahedra.

Considering the equal contributions of $D$ and $E$, we conclude
that this picture of an ordering of $d_{3z^2-r^2}$ orbitals has to
be modified as it results in $E=0$ in contradiction to our
findings, which are also in agreement with the results of Pimenov
\textit{et al.} \cite{Pimenov00}. In a local coordinate system the
orbital of the $e_g$ electron is a superposition
$\psi_{g,e}=c_1\phi_{3z^2-r^2}\pm c_2\phi_{x^2-y^2}$
\cite{Matsumoto70}, where $c_1 = 0.8$ and $c_2 = 0.6$ denote the
orbital mixing coefficients down to lowest temperatures as
reported by Rodriguez-Carvajal \textit{et
al.}~\cite{Rodriguez-Carvajal98} on the base of
neutron-diffraction studies. Matsumoto \cite{Matsumoto70} found in
second order perturbation theory the following formula for the CF
parameter $E$

\begin{equation}
E=-2\sqrt{3}\frac{\lambda^2}{\Delta}c_1c_2,
\end{equation}

with the spin-orbit coupling $\lambda$ and the energy separation
$\Delta$ between the $^5E_g$ ground state and $^3T_{1g}$. This
formula qualitatively confirms our conclusions, as not only the
sign of our parameters is reproduced but also both orbital mixing
coefficients must not be negligibly small in order to explain the
observed rhombic CF parameter E.

\section{CONCLUSION}

We performed a systematic investigation of the angular dependence
of the paramagnetic resonance in the Jahn-Teller distorted
orthorhombic phase in La$_{0.95}$Sr$_{0.05}$MnO$_3$ single
crystals. We presented a comprehensive analysis for the resonance
linewidth in high-temperature approximation, which takes into
account the microscopic geometry of the four inequivalent Mn
positions in the orthorhombic unit cell based on the structural
data determined for LaMnO$_3$ by Huang \textit {et
al.}~\cite{huang}. The crystal-field parameters for all Mn
positions and the Dzyaloshinsky-Moriya interaction for
nearest-neighbor Mn ions along the $b$ axis as well as in the $ac$
plane were successfully extracted as $D(\infty) = 0.91$~K,
$E(\infty) = -0.79$~K, $d_1(\infty) = 1.26$~K and $d_2(\infty) =
0.38$~K. These findings shed new light on the microscopic picture
of orbital ordering and spin-spin interaction in these compounds.

\section*{ACKNOWLEDGEMENTS}

It is a pleasure to thank K.-H.~H\"{o}ck for fruitful discussions.
This work was supported in part by the BMBF under contract
no.~13N6917 (EKM), by the Deutsche Forschungsgemeinschaft (DFG)
via the Sonderforschungsbereich 484 and DFG-project No.~436-RUS
113/566/0, and by INTAS (project no.~97-30850). The work of
M.~V.~Eremin was partially supported by RFBR grant no.~00-02-17597
and NIOKR Tatarstan.

\section*{APPENDIX 1: Relative superexchange strength and sign of parameters}

We consider the Hamiltonian of superexchange coupling between two
Mn$^{3+}$ ions on lattice sites $(A)$ and $(B)$

\begin{equation}
{\cal H}=J_{AB}({\mathbf S}_{A} \cdot {\mathbf
S}_{B})=2\sum\limits_{\eta ,\zeta }j_{\eta \zeta ,\zeta \eta
}({\mathbf s}_{\eta }\cdot{\mathbf s}_{\zeta})
\end{equation}

where $j_{\eta \zeta ^{\prime },\zeta \eta ^{\prime }}$ denote the
superexchange parameters via one electron states, the symbol
$\eta $ refers to the Mn($A$) state whereas $\zeta$ refers to
Mn($B$). In the ground state Mn$^{3+}$ ions have maximum spin.
Therefore we can write:

\begin{equation}
J_{AB}=\frac{1}{2S_{A}S_{B}}\sum\limits_{\eta ,\zeta }j_{\eta
\zeta ,\zeta \eta }
\end{equation}

Along the $b$ axis, the superexchange is mainly realized via the
$|y^{2}-x^{2}\rangle-| y^{2}-x^{2}\rangle$ channel and has
antiferromagnetic character. Within the $ac$ plane the
ferromagnetic coupling is mainly transferred via the channel
$|y^{2}-x^{2}\rangle-|y^{2}-z^{2}\rangle$. The relevant states
like $| y^{2}-x^{2}\rangle$ and $| y^{2}-z^{2}\rangle$ are
mutually orthogonal and yield negative
$j_{y^{2}-x^{2},y^{2}-z^{2}}$.

The Dzyaloshinsky-Moriya parameter $d_{AB}$ from equation
\ref{dmham} can be derived from the following expression

\begin{eqnarray}
G_{AB}^{\perp}&=&-\frac{i}{2S_{A}S_{B}}\{\sum\limits_{\eta ,\zeta
,\eta^{\prime }} \frac{\xi_{A}}{| \Delta_{\eta ,\eta ^{\prime }}|}
j_{\eta \zeta ,\zeta \eta ^{\prime }}\langle\eta
^{\prime }|l_{A}^{(\perp)}| \eta \rangle \nonumber \\
&-& \sum\limits_{\eta ,\zeta ,\eta ^{\prime }}\frac{\xi _{B}}{ |
\Delta_{\zeta ,\zeta ^{\prime }}| }j_{\eta \zeta ^{\prime },\zeta
\eta }\langle\zeta ^{\prime }| l_{B}^{(\perp )}| \zeta \rangle\}
\end{eqnarray}

here $i$ is the imaginary unit, $\xi _{A}$ and $\xi _{B}$ are
spin-orbit coupling constants, $\Delta_{\zeta ,\zeta ^{\prime }}$
denotes the energy splitting between the states $\zeta$ and $\zeta
^{\prime }$. $l_{A}^{(\perp )}$ and $l_{B}^{(\perp )}$ are
components of the one-electron orbital momentum perpendicular to
the plane, built up by Mn($A$)-bridging oxygen(O) -Mn($B$),
${\mathbf n}_{AO}$ and ${\mathbf n}_{BO}$ are the corresponding
unit vectors along the Mn-O bonds. $j_{\eta \zeta^{\prime} ,\zeta
\eta }$ and $j_{\eta \zeta ,\zeta \eta ^{\prime }}$ can be
expressed via a product of dimensionless transfer integrals
$\lambda_{\sigma },\lambda_{\pi},\lambda_{s}$ corresponding to
Mn-O bonds \cite{anderson2,eremin}

In order to explain the main features for the moment, we shall
simplify the orbital ordering as $| y^{2}-z^{2}\rangle$ like
states along the $b$ axis and alternating ordering like $|
y^{2}-z^{2}\rangle$, $| y^{2}-x^{2}\rangle$ within the $(ac)$
plane. Then one has the following situation:

1) Along the $b$ axis, the largest contribution to the
superexchange comes from
$j_{y^{2}-z^{2},y^{2}-z^{2},y^{2}-z^{2},y^{2}-z^{2}}$, yielding

\begin{equation}
J_{b}\sim \frac{1}{8}(\frac{3}{4})^{2}[\lambda _{\sigma }^{2}\cos
\vartheta +\lambda _{s}^{2}]^{2}
\end{equation}

and for the DM vector the most important term is
$j_{y^{2}-z^{2},y^{2}-z^{2},y^{2}-z^{2},yz}$, which leads to

\begin{equation}
d_{1}\sim -\frac{\xi }{2\Delta }(\frac{3}{4})^{3/2}[\lambda
_{\sigma }^{2}\cos \vartheta +\lambda _{s}^{2}]\lambda _{\pi
}\lambda _{\sigma }
\end{equation}

2) Within the $ac$ plane, the major contribution to the
superexchange comes from
$j_{y^{2}-z^{2},y^{2}-x^{2},y^{2}-x^{2},y^{2}-z^{2}}$ and hence

\begin{equation}
J_{ac}\sim -\frac{J_{\rm H}}{8U}(\frac{3}{4})[\lambda _{\sigma
}^{2}\cos \vartheta +\lambda _{s}^{2}]^{2}
\end{equation}

whereas for $\ DM$ vector the most important term is
$j_{y^{2}-z^{2},y^{2}-z^{2},y^{2}-z^{2},xy}$ and therefore

\begin{equation}
d_{2}\sim \frac{J_{\rm H}}{4U}\frac{\xi }{\Delta
}(\frac{3}{4})^{1/2}[\lambda _{\sigma }^{2}\cos \vartheta
+\lambda _{s}^{2}]\lambda _{\pi }\lambda _{\sigma }
\end{equation}

here $\vartheta$ is the angle between ${\mathbf n}_{AO}$ and
${\mathbf n}_{OB}$ vectors, $J_{\rm H}$ denotes the intra-atomic
Hund's exchange parameter, $U$ is the charge-transfer energy
between the Mn ions. From these expression one can clearly see
that the relative sign of parameters $d_{AB}$ along the $b$ axis
($d_{1}$ in text) and $d_{AB}$ within $(ac)$ plane ($d_{2}$ in
text) are different as well as superexchange parameters $J_{b}$
and $J_{ac}$.

\section*{APPENDIX 2: remarks on the calculation}

Here we outline the transformation between the crystallographic
system $(x,y,z)||(a,b,c)$ and the coordinate system $(\tilde{x},
\tilde{y}, \tilde{z})$ with the $\tilde{z}$-axis parallel to the
applied magnetic field $H$ and rotated by the polar angle $\theta$
and the azimuth angle $\varphi$ with respect to the
crystallographic system. Assuming an isotropic $g$ value, the DM
vector transforms like the space coordinates:

\begin{eqnarray}
\tilde{G}_{ij}^{x}&=& G_{ij}^{x}\cos \theta \cos \varphi
+G_{ij}^{y}\cos \theta \sin \varphi -G_{ij}^{z}\sin \theta
\nonumber
\\ \tilde{G}_{ij}^{y}&=& G_{ij}^{y}\cos \varphi -G_{ij}^{x}\sin \varphi
\nonumber \\ \tilde{G}_{ij}^{z}&=& G_{ij}^{x}\sin \theta \cos
\varphi +G_{ij}^{y}\sin \theta \sin \varphi +G_{ij}^{z}\cos \theta
\end{eqnarray}

The crystal-field components $D_{\alpha \beta} \propto (\alpha
\beta)/R^2$ transform like the product of the oxygen coordinates
$\alpha$ and $\beta$.

For the angular dependence of the resonance field we need the
following relations:

1) ab plane $(\theta =\frac{\pi }{2})$

\begin{eqnarray}
\sum\limits_j
(\tilde{D}_{xx}^{(j)}+\tilde{D}_{yy}^{(j)}-2\tilde{D}_{zz}^{(j)})
&=& \frac{1}{2}\sum\limits_i(2D_{zz}^{(i)}-D_{xx}^{(i)}-
D_{yy}^{(i)})\nonumber
\\ &-& \frac{3}{2} \sum\limits_i
(D_{xx}^{(i)}-D_{yy}^{(i)})\cos 2\varphi \nonumber \\
\sum\limits_j (\tilde{D}_{xx}^{(j)}-\tilde{D}_{yy}^{(j)} \pm
2i\tilde{D}_{xy}^{(j)})
&=&\frac{1}{2}\sum\limits_i(2D_{zz}^{(i)}-D_{xx}^{(i)}-D_{yy}^{(i)})\nonumber
\\ &+& \frac{1}{2} \sum\limits_i (D_{xx}^{(i)}-D_{yy}^{(i)})\cos
2\varphi \nonumber
\end{eqnarray}

2) ac plane $(\varphi =0)$

\begin{eqnarray}
\sum\limits_j
(\tilde{D}_{xx}^{(j)}+\tilde{D}_{yy}^{(j)}-2\tilde{D}_{zz}^{(j)})
&=&\frac{1}{2}\sum\limits_i
(2D_{yy}^{(i)}-D_{zz}^{(i)}-D_{xx}^{(i)})\nonumber
\\&-&\frac{3}{2}\sum\limits_i
(D_{zz}^{(i)}-D_{xx}^{(i)})\cos 2\theta \nonumber \\
\sum\limits_j (\tilde{D}_{xx}^{(j)}-\tilde{D}_{yy}^{(j)} \pm
2i\tilde{D}_{xy}^{(j)})
&=&-\frac{1}{2}\sum\limits_i(2D_{yy}^{(i)}-D_{zz}^{(i)}-D_{xx}^{(i)})
\nonumber \\&-&\frac{1}{2}\sum\limits_i
(D_{zz}^{(i)}-D_{xx}^{(i)})\cos 2\theta \nonumber
\end{eqnarray}

3) bc plane $(\varphi =\frac{\pi }{2})$

\begin{eqnarray}
\sum\limits_j(\tilde{D}_{xx}^{(j)}+\tilde{D}_{yy}^{(j)}-2\tilde{D}_{zz}^{(j)})
&=&\frac{1}{2}\sum\limits_i(2D_{xx}^{(i)}-D_{yy}^{(i)}-D_{zz}^{(i)})\nonumber
\\&-& \frac{3}{2}\sum\limits_i(D_{zz}^{(i)}-D_{yy}^{(i)})\cos
2\theta \nonumber
\\
\sum\limits_j (\tilde{D}_{xx}^{(j)}-\tilde{D}_{yy}^{(j)} \pm
2i\tilde{D}_{xy}^{(j)})
&=&-\frac{1}{2}\sum\limits_i(2D_{xx}^{(i)}-D_{yy}^{(i)}-D_{zz}^{(i)})
\nonumber \\&-&\frac{1}{2}\sum\limits_i
(D_{zz}^{(i)}-D_{yy}^{(i)})\cos 2\theta \nonumber
\end{eqnarray}

Concerning the linewidth, the second moment due to DM interaction
can be expressed in crystallographic coordinates as

\begin{eqnarray}
M_{2}^{\rm DM} &=& \frac{2}{3}S(S+1)\{
\sum\limits_{i,j}(G_{ij}^{x})^{2}[1+\sin^{2}\theta\cos^{2}\varphi]\nonumber
\\
&+&\sum\limits_{i,j}(G_{ij}^{y})^{2}[1+\sin^{2}\theta\sin^{2}\varphi]\\
&+& \sum\limits_{i,j}(G_{ij}^{z})^{2}[1+\cos^{2}\theta]
+\sum\limits_{i,j}G_{ij}^{x}G_{ij}^{y}\sin 2\varphi\sin^{2}\theta
\nonumber \\ &+& \sum\limits_{i,j}G_{ij}^{x}G_{ij}^{z}\sin 2\theta
\cos \varphi +\sum\limits_{i,j}G_{ij}^{y}G_{ij}^{z}\sin 2\theta
\sin \varphi \}\nonumber
\end{eqnarray}

On evaluation of $M_{2}^{\rm DM}$, it turns out that all sums over
cross terms $G_{ij}^{\alpha}G_{ij}^{\beta}$ with $\alpha \neq
\beta$ vanish and the remaining sums over $i$, which are equal
for all Mn sites $j$, read:

\begin{eqnarray}
\sum\limits_{i}(G_{ij}^{x})^{2}=2\alpha_{b}^{2}+4\alpha_{ab}^{2}
\nonumber \\
\sum\limits_{i}(G_{ij}^{y})^{2}=4\beta _{ab}^{2} \nonumber \\
\sum\limits_{i}(G_{ij}^{z})^{2}=2\gamma_{b}^{2}+4\gamma_{ab}^{2}
\end{eqnarray}

Concerning the contribution of the crystal field $M_{2}^{\rm CF}$,
it is clear from values listed in Table \ref{CFpara} that many
sums over four Mn position vanish in the crystallographic system,
which is very convenient. For example

\begin{eqnarray}
\sum\limits_j
(D_{zz}^{(j)}-D_{yy}^{(j)})D_{yz}^{(j)}&=&\sum\limits_j
(D_{zz}^{(j)}-D_{xx}^{(j)})D_{xz}^{(j)}\nonumber \\
&=&\sum\limits_j
(D_{xx}^{(j)}-D_{yy}^{(j)})D_{xy}^{(j)}\nonumber \\
&=&\sum\limits_j
[2D_{xx}^{(j)}-D_{zz}^{(j)}-D_{yy}^{(j)}]D_{yz}^{(j)} \nonumber \\
&=& \sum\limits_j
[2D_{yy}^{(j)}-D_{xx}^{(j)}-D_{zz}^{(j)}]D_{xz}^{(j)} \nonumber \\
&=&\sum\limits_j
[2D_{zz}^{(j)}-D_{xx}^{(j)}-D_{yy}^{(j)}]D_{xy}^{(j)}=0 \nonumber
\end{eqnarray}

From symmetry point of view it is understandable, because there
are mirror planes within $ab$, $ac$ and $bc$ planes. Having in
mind these properties, after some manipulation we arrive to the
following formulae for the secular part of $M_{2}^{\rm CF}$
(i.e.~when factors $f_{2}$ and $f_{3}$ are equal to zero) and
nonsecular parts (we consider the case $f_{2}$ = $f_{3}$):

1) $ab$ plane $(\theta =\frac{\pi }{2})$:

\begin{eqnarray}
M_{2}^{\rm CF}({\rm sec})
&=&\frac{1}{80}[4S(S+1)-3]\{\frac{1}{4}\sum\limits_j
[2D_{zz}^{(j)}-D_{xx}^{(j)}-D_{yy}^{(j)}]^{2} \nonumber \\
&+&\frac{9}{8}\sum\limits_j[(D_{xx}^{(j)}-D_{yy}^{(j)})^{2}+4(D_{xy}^{(j)})^{2}]
\nonumber \\
&-&\frac{3}{2}\sum\limits_j[2D_{zz}^{(j)}-D_{xx}^{(j)}-D_{yy}^{(j)}]
(D_{xx}^{(j)}-D_{yy}^{(j)})\cos2\varphi \nonumber \\
&+&\frac{9}{8}\sum\limits_j[(D_{xx}^{(j)}-D_{yy}^{(j)})^{2}-4(D_{xy}^{(j)})^{2}]
\cos4\varphi \}
\end{eqnarray}

\begin{eqnarray}
M_{2}^{\rm CF}({\rm non})
&=&\frac{1}{80}[4S(S+1)-3]\{\frac{1}{4}\sum\limits_j
[2D_{zz}^{(j)}-D_{yy}^{(j)}-D_{xx}^{(j)}]^{2} \nonumber \\
&+&\frac{11}{8}\sum\limits_j
[(D_{yy}^{(j)}-D_{xx}^{(j)})^{2}+4(D_{xy}^{(j)})^{2}] \nonumber
\\ &+& 7\sum\limits_j[(D_{zy}^{(j)})^{2}+(D_{xz}^{(j)})^{2}] \nonumber \\
&+&\sum\limits_j
\{\frac{1}{2}[2D_{zz}^{(j)}-D_{yy}^{(j)}-D_{xx}^{(j)}
](D_{xx}^{(j)}-D_{yy}^{(j)})\nonumber
\\&+& 3[(D_{zy}^{(j)})^{2}-(D_{xz}^{(j)})^{2}]\}
\cos2\varphi \nonumber \\
&-&\frac{9}{8}\sum\limits_j
[(D_{xx}^{(j)}-D_{yy}^{(j)})^{2}-4(D_{xy}^{(j)})^{2}]\cos
4\varphi \}
\end{eqnarray}

2) $ac$ plane $(\varphi =0)$: like $ab$ plane with permutation
${x,y,z}\rightarrow{z,x,y}$ and exchange
$\varphi\rightarrow\theta$.

3) $bc$ plane $(\varphi =\frac{\pi }{2})$: like $ab$ plane with
permutation ${x,y,z}\rightarrow{z,y,x}$ and exchange
$\varphi\rightarrow\theta$.

\clearpage
\begin{table}
\caption[]{Crystal field parameters for different Mn positions in
the unit cell.} \label{CFpara}
\begin{tabular}{lllll}
$D_{\alpha \beta }\backslash $Mn (j)- site & $(0,0,\frac{1}{2})$ & $(\frac{1%
}{2},0,1)$ & $(0,\frac{1}{2},\frac{1}{2})$ &
$(\frac{1}{2},\frac{1}{2},1)$
\\
$D_{xx}^{(j)}$ & $D_{xx}^{(1)}$ & $D_{xx}^{(1)}$ & $D_{xx}^{(1)}$ & $%
D_{xx}^{(1)}$ \\
$D_{yy}^{(j)}$ & $D_{yy}^{(1)}$ & $D_{yy}^{(1)}$ & $D_{yy}^{(1)}$ & $%
D_{yy}^{(1)}$ \\
$D_{zz}^{(j)}$ & $D_{zz}^{(1)}$ & $D_{zz}^{(1)}$ & $D_{zz}^{(1)}$ & $%
D_{zz}^{(1)}$ \\
$D_{xy}^{(j)}$ & $D_{xy}^{(1)}$ & $D_{xy}^{(1)}$ & $-D_{xy}^{(1)}$ & $%
-D_{xy}^{(1)}$ \\
$D_{xz}^{(j)}$ & $D_{xz}^{(1)}$ & $-D_{xz}^{(1)}$ & $D_{xz}^{(1)}$ & $%
-D_{xz}^{(1)}$ \\
$D_{yz}^{(j)}$ & $D_{yz}^{(1)}$ & $-D_{yz}^{(1)}$ & $-D_{yz}^{(1)}$ & $%
D_{yz}^{(1)}$%
\end{tabular}
\end{table}

\begin{table}
\caption[]{DM vector components for different pairs of Mn ions in
the unit cell of LaMnO$_{3}$.} \label{DMpara}
\begin{tabular}{lllll}
Mn(i)-site$\setminus $Mn (j)-site & $(0,0,\frac{1}{2})$ & $(\frac{1}{2}%
,0,1)$ & $(0,\frac{1}{2},\frac{1}{2})$ & $(\frac{1}{2},\frac{1}{2},1)$ \\
$(x_{j},y_{j}+\frac{1}{2},z_{j})$ & $(-\alpha _{b},0,-\gamma _{b})$ & $%
(\alpha _{b},0,-\gamma _{b})$ & $(-\alpha _{b},0,-\gamma _{b})$ &
$(\alpha
_{b},0,-\gamma _{b})$ \\
$(x_{j},y_{j}-\frac{1}{2},z_{j})$ & $(\alpha _{b},0,\gamma _{b})$ & $%
(-\alpha _{b},0,\gamma _{b})$ & $(\alpha _{b},0,\gamma _{b})$ &
$(-\alpha
_{b},0,\gamma _{b})$ \\
$(x_{j}+\frac{1}{2},y_{j},z_{j}-\frac{1}{2})$ & $(-\alpha
_{ac},\beta
_{ac},-\gamma _{ac})$ & $(\alpha _{ac},-\beta _{ac},\gamma _{ac})$ & $%
(\alpha _{ac},\beta _{ac},\gamma _{ac})$ & $(-\alpha _{ac},-\beta
_{ac},-\gamma _{ac})$ \\
$(x_{j}-\frac{1}{2},y_{j},z_{j}-\frac{1}{2})$ & $(-\alpha
_{ac},\beta
_{ac},\gamma _{ac})$ & $(\alpha _{ac},-\beta _{ac},-\gamma _{ac})$ & $%
(\alpha _{ac},\beta _{ac},-\gamma _{ac})$ & $(-\alpha _{ac},-\beta
_{ac},\gamma _{ac})$ \\
$(x_{j}-\frac{1}{2},y_{j},z_{j}+\frac{1}{2})$ & $(-\alpha
_{ac},\beta
_{ac},-\gamma _{ac})$ & $(\alpha _{ac},-\beta _{ac},\gamma _{ac})$ & $%
(\alpha _{ac},\beta _{ac},\gamma _{ac})$ & $(-\alpha _{ac},-\beta
_{ac},-\gamma _{ac})$ \\
$(x_{j}+\frac{1}{2},y_{j},z_{j}+\frac{1}{2})$ & $(-\alpha
_{ac},\beta
_{ac},\gamma _{ac})$ & $(\alpha _{ac},-\beta _{ac},-\gamma _{ac})$ & $%
(\alpha _{ac},\beta _{ac},-\gamma _{ac})$ & $(-\alpha _{ac},-\beta
_{ac},\gamma _{ac})$%
\end{tabular}
\end{table}

\end{document}